\documentclass[fleqn,usenatbib]{mnras}

\usepackage{newtxtext,newtxmath}

\usepackage[T1]{fontenc}

\DeclareRobustCommand{\VAN}[3]{#2}
\let\VANthebibliography\thebibliography
\def\thebibliography{\DeclareRobustCommand{\VAN}[3]{##3}\VANthebibliography}

\usepackage{graphicx}	
\usepackage{amsmath}	
\usepackage{multicol}        
\usepackage{bm}		
\usepackage{pdflscape}	
\usepackage{longtable}  
\usepackage{caption} 
\usepackage{subfigure} 
\usepackage[flushleft]{threeparttable}
\usepackage{multirow}   
\usepackage{xfrac}   
\usepackage{xcolor} 
\usepackage{todonotes} 
\usepackage{outlines} 
\usepackage{ulem} 
\usepackage{subfigure}

\newcommand{\ignore}[1]{}

\title[MeerTRAP transients]{Discovery of 27 new Rotating Radio Transients by MeerTRAP}

\author[S. Singh et al.]{S. Singh$^{1}$\thanks{E-mail: shubham.singh@manchester.ac.uk}, J. Tian$^{1}$\thanks{E-mail: jun.tian@manchester.ac.uk}, B. W. Stappers$^1$, K. Shaji$^{5,6}$, K. M. Rajwade$^2$, L. Levin$^{1}$, J. D. Turner$^{8}$,\newauthor M. C. Bezuidenhout$^{3,4}$, 
 M. Caleb$^{5,6}$, I. Pastor-Marazuela$^{7,1}$, F. Jankowski$^{14}$, R. Karuppusamy$^{9}$, \newauthor E. D. Barr$^{9}$, M. Kramer$^{9}$, R. Breton$^{1}$, C. J. Clark$^{10,11}$, T. Thongmeearkom$^{1,12}$, M. Burgay$^{13}$
 \\\\
$^{1}$Jodrell Bank Centre for Astrophysics, Department of Physics and Astronomy, The University of Manchester, Manchester M13 9PL, UK\\
$^2$Astrophysics, The University of Oxford, Denys Wilkinson Building, Keble Road,
Oxford OX1 3RH, UK\\
$^3$SKA Observatory, 2 Fir Street, Observatory 7925, Cape Town, South Africa\\
$^{4}$South African Radio Astronomy Observatory, 2 Fir Street, Observatory
7925, Cape Town, South Africa\\
$^5$Sydney Institute for Astronomy, School of Physics, The University of Sydney, NSW 2006, Australia\\
$^6$ARC Centre of Excellence for Gravitational Wave Discovery (OzGrav), Hawthorn, 3122, Victoria, Australia\\
$^{7}$ ASTRON, the Netherlands Institute for Radio Astronomy, Oude Hoogeveensedijk 4,7991 PD Dwingeloo, The Netherlands\\
$^{8}$INAF - Osservatorio Astronomico di Roma, Via Frascati 33, I-00078 Monte Porzio Catone, Italy\\
$^{9}$Max-Planck-Institut fur Radioastronomie, 53121 Bonn, Germany\\
$^{10}$Max Planck Institute for Gravitational Physics (Albert Einstein Institute), D-30167 Hannover, Germany\\
$^{11}$Leibniz Universität Hannover, D-30167 Hannover, Germany\\
$^{12}$National Astronomical Research Institute of Thailand, Don Kaeo, Mae Rim, Chiang Mai 50180, Thailand\\
$^{13}$INAF-Osservatorio Astronomico di Cagliari, Via della Scienza 5, 09047 Selargius (CA), Italy\\
$^{14}$LPC2E, OSUC, Univ Orleans, CNRS, CNES, Observatoire de Paris, F-45071 Orleans, France\\
}

\date{Accepted XXX. Received YYY; in original form ZZZ}

\pubyear{2026}

\begin{document}
\label{firstpage}
\pagerange{\pageref{firstpage}--\pageref{lastpage}}

\maketitle

\begin{abstract}
We present the discovery of 27 new Galactic transients made by the commensal MeerTRAP single pulse search programme at the MeerKAT telescope. We determine the location of 14 of these discoveries with arcsecond accuracy by imaging the data captured in the dedicated transient buffer. A preliminary estimate of the period was made for 8 sources using the arrival times of several pulses detected by MeerTRAP. The periods of these transients range between 0.78 s and 4 s. For the previously published MeerTRAP source MTP0040 (PSR J1357$-$6507), we are able to provide the position and period using new pulses detected since it was reported. We also estimate the radio fluences and discuss the burst rates derived from the commensal search detections. The accurate image-domain localizations were used to conduct follow-up observations and timing analysis. Furthermore, some follow-up observations revealed ordinary pulsar-like pulsed behavior for four of the new sources, illustrating the blurred separation between rotating radio transients and canonical pulsars. We also present coherent timing solutions for 4 sources, providing insight into their rotational properties and their place within the Galactic neutron star population.
\end{abstract}




\section{Introduction}\label{intro}
Pulsars are rapidly rotating, highly magnetized neutron stars that are best known for their pulsed radio emission. To date, more than 4000 pulsars have been discovered \citep{ATNF}. While the majority of pulsars emit consistently at radio wavelengths, a small subset exhibits puzzling single-pulse behaviors, e.g., intermittent emission or nulling \citep{Backer70, Biggs92, Wang07}, giant pulse emission \citep{giant_pulses_staelin, Cognard96, Knight06, Mickaliger12} and microstructure \citep{Hankins71, Cordes90, Kramer02}. Understanding these diverse phenomena is the key to establishing a unifying model for the neutron star zoo \citep{NS_birthrate_problem, Kaspi10}.

Rotating radio transients (RRATs) are a subclass of pulsars that exhibit occasionally observable emission in the form of single radio pulses \citep{McLaughlin06}. Timing analysis confirmed they are indeed periodic rotators \citep{McLaughlin09, RRAT_review}. This intermittent emission from RRATs is more detectable via single pulse searches than periodicity searches \citep{McLaughlin03}. The RRAT population has rapidly expanded in recent years thanks to the growing popularity of single pulse searches \citep{Han21, Good21, Bezuidenhout22, FAST_RRATs, Dong23, MeerTRAP_james, Han25, MeerTRAPIII}. Despite extensive survey efforts, only $\sim336$ RRATs have been discovered, and even fewer have timing solutions (see RRATalog\footnote{\url{https://rratalog.github.io/rratalog/}}, \citealt{RRATlog_ref}), yet they are thought to be as populous as canonical radio pulsars \citep{McLaughlin06, RRAT_review}. In the $\text{P}-\dot{\text{P}}$ diagram, these RRATs typically have longer periods and higher magnetic fields than normal pulsars \citep{Burke10, Karako15, Cui17}. However, it remains unclear whether these are intrinsic properties of RRATs or a result of observational bias. Finding more RRATs and measuring their period and period derivative will help us gain a more complete understanding of the true pulsar population and its evolutionary paths.

Magnetars are another subclass of neutron stars; they are magnetically powered and feature a wide range of X-ray activities (for a review see \citealt{magnetar_review_kaspi}). Some magnetars exhibit sporadic radio emission similar to that observed for RRATs \citep{Camilo06}. Some RRATs are known to exhibit magnetar-like X-ray outbursts (e.g., \citealt{Rea09, Archibald17}). While many RRATs are situated close to the magnetars in $\text{P}-\dot{\text{P}}$ diagram \citep{Cui17}, studies over the last decade question any direct evolutionary link between the two classes \citep{Gencali24, NS_evolution2013}. 
 
Recently, a new class of rotating neutron stars, long-period transients (LPTs\footnote{Note that some LPTs are likely to be white dwarf binaries, e.g. GLEAM-X J0704$-$37 \citep{Hurley24} and ILT J1101+5521 \citep{Ruiter25}.}), have been discovered at radio wavelengths. Their periods range from a few tens of seconds to a few hours \citep{76s_pulsar, Caleb24, Hurley22, Hurley23, Wang24, McSweeney25, Dong25a, Dong25b, Anumarlapudi25}. These sources challenge the conventional understanding of coherent radio emission from neutron stars. Some LPTs also show X-ray emission \citep{Wang24, Anumarlapudi25}. It has been suggested that some LPTs could be radio-loud magnetars \citep{76s_pulsar, Lee25}. Only a small number of pulsars have been discovered with periods between a few tens of seconds and a few minutes \citep{Tan18, Morello20, 76s_pulsar, Surnis23, Wang24, Dong25b}. As Galactic transients such as RRATs and magnetars are mostly long-period sources, finding more of them will fill in this parameter space, inform us about the distribution of different populations of neutron stars, and potentially establish an evolutionary link between them.

To date, many pulsar and RRAT surveys have been carried out, either with interferometers featuring wide fields of view, e.g. the Low-Frequency Array (LOFAR; \citealt{Sanidas19}), Canadian Hydrogen Mapping Experiment (CHIME; \citealt{CHIME/Pulsar21}), Murchison Widefield Array (MWA; \citealt{Bhat23a}) and Australian SKA Pathfinder (ASKAP; \citealt{Shannon25, Wang25b}), or sensitive single-dish telescopes, e.g. Effelsberg \citep{Barr13}, Murriyang \citep{Keith10} and the Five-hundred-meter Aperture Spherical radio Telescope (FAST; \citealt{Han21}). Remarkably, the FAST Galactic Plane Pulsar Snapshot survey recently discovered 107 RRATs \citep{Han21, FAST_RRATs, Han25}, and the Meer(more) TRAnsients and Pulsars (MeerTRAP; \citealt{stappers2016}) survey at MeerKAT discovered 68 RRATs \citep{Bezuidenhout22, MeerTRAP_james, MeerTRAPIII}. These surveys alone almost doubled the known RRAT population. However, only a small fraction of RRATs have a period measurement, and even fewer have a measurement of the period derivative. Dedicated follow-up observations are therefore essential to characterize their rotational properties and determine their relationship to the broader neutron star population.

In this paper, we present the discovery of 27 RRATs by MeerKAT and subsequent follow-up with the 100-m Effelsberg telescope and the 64-m Murriyang telescope at Parkes. In \autoref{sec:obs}, we briefly describe the observations and analysis process. We present the new discoveries and analysis results in \autoref{sec:results}. We then discuss the results and their implications in the wider context of transient searches in \autoref{sec:disc}.

\section{Observations and Data Reduction}\label{sec:obs}
\subsection{MeerTRAP}\label{sec:MeerTRAP}
MeerTRAP is a commensal search programme for fast radio transients with the MeerKAT telescope \citep{stappers2016}. It can simultaneously form one incoherent beam (IB) and up to 768 coherent beams (CBs) on the sky that cover a portion of the field of view of MeerKAT. Each of the CBs has a half-power width of $\sim1$\,arcmin. The beamformed data are received from FBFUSE \citep{FBFUSE_barr} at a time / frequency resolution of 0.53\,MHz / $481.88\,\mu$s, 0.84\,MHz / $306.24\,\mu$s and 0.85\,MHz / $449.39\,\mu$s for MeerKAT UHF (544--1088\,MHz), L-band (856--1712\,MHz) and S-band (1968--2843\,MHz), respectively. These data are searched for dispersed single pulses in real time. We excise broadband radio frequency interference (RFI) using the zero-DM technique \citep{ZDMF_2019, Eatough09} and the frequency channels dominated by narrowband RFI were masked using the {\sc iqrm} algorithm \citep{Morello22} with standard deviation as the metric and a 3 sigma threshold for outlier detection. Any detections with a dispersion measure (DM) below $20\,\text{pc}\,\text{cm}^{-3}$ are rejected to further remove broadband RFI. Candidates with a signal-to-noise ratio (S/N) above 8 are sifted and classified using {\sc frbid}\footnote{\url{https://github.com/Zafiirah13/FRBID}} before being manually inspected. More details about the MeerTRAP search pipeline can be found in \citet{Rajwade22}.

MeerTRAP piggybacks on other MeerKAT observations and thus is likely to detect sporadic single pulses from pulsars and RRATs. We save $\sim1$\,s of total intensity data for every detected pulse along with the detection DM and the coordinates of the detection beam. We compare the detection DM with the \texttt{NE2001} \citep{NE2001} and \texttt{YMW16} \citep{YMW16} Galactic electron density models to distinguish Galactic and extragalactic transients. MeerKAT observations can spend hours on a single target, providing MeerTRAP with an opportunity to detect pulses from all types of transients in the field of view, including Galactic radio transients with extremely low burst rates. If several pulses are detected from a source, we can solve for its period. More details about this period search method can be found in \citet{MeerTRAP_james} and \citet{MeerTRAPIII}. When the same, or other, MeerKAT projects revisit the same part of the sky multiple times, more detections of these transients over several months provide a chance to obtain a coherent timing solution.

Raw voltage data are saved for some transients that trigger the transient buffer. These data are correlated with \texttt{xGPU} \citep{clark_accelerating_2011} and then converted to visibilities with \texttt{DifX} \citep{deller_difx_2007, deller_difx-2_2011}. We make images of the transient using {\sc wsclean} \citep{Offringa14} and identify it in the detection beam, allowing for localization to arcsecond precision. More details about the voltage data reduction and the localization method can be found in \citet{Rajwade24}.


\subsection{Effelsberg and Murriyang follow-up}
We conducted follow-up observations of some of the MeerTRAP discoveries using other radio telescopes. The criteria for deciding which sources to prioritize for these observations included good localization accuracy, higher fluence and burst rate, and/or interesting pulse morphology. This resulted in follow up of three MeerTRAP-discovered sources with the 100-metre Effelsberg radio telescope \citep{Effelseberg_paper} to search for periodic signals as well as single pulses. These observations were carried out with the Ultra Broad-Band (UBB) receiver that covers a wide frequency range from 1.1\,GHz to 6\,GHz. The observations were coherently dedispersed to the discovery DM of the target transient. The observation of each source lasted roughly for 2 hours. The full band was evenly split into seven subbands, and the data were recorded in the search mode with 8-bit full Stokes and $128\,\mu$s time resolution and 0.58\,MHz frequency resolution.

We followed up another eight MeerTRAP sources with the Murriyang telescope at Parkes Observatory, using the ultrawide-band low (UWL) receiver \citep{Hobbs20}. Each source was observed continuously for $\sim2$ hours, and the data were recorded with a time resolution of $256\,\mu$s and a frequency resolution of 1\,MHz after coherent dedispersion at the discovery DM. We split the bandwidth between 704\,MHz and 4\,GHz evenly into eight subbands. A 2\,min scan of a noise diode was performed before each observation to be used for calibration.

We searched the subbanded data for single pulses using {\sc transientx}\footnote{\url{https://github.com/ypmen/TransientX}} \citep{TransientX}. {\sc filtool} \citep{pulsarX} was used for RFI cleaning. We searched in a DM range of $\pm10\%$ around the discovery DM with a step size of $0.1\,\text{pc}\,\text{cm}^{-3}$ and a maximum boxcar width of 100\, ms. We manually inspected all candidates with $\text{S/N}>8$ to identify real signals.

Finding periodic signals from the Effelsberg and Parkes observations would allow us to confirm the spin periods of the MeerTRAP sources and study their folded pulse properties. We searched the follow-up observations for faint periodic emission using the Fast Folding Algorithm (FFA). The FFA search is expected to be sensitive to sources with long periods, small duty cycles \citep{singh_FFA, FFA_morello}, and sources with high nulling \citep{GHRSS_VI, grover_FFA}, and thus is ideal to find any faint counterpart of RRAT-like emission. We used the {\sc rseek} utility of the {\sc riptide}\footnote{\url{https://github.com/v-morello/riptide}} package \citep{FFA_morello}. We searched in the period range of 0.1 s to 100 s and duty cycle range of 0.2\% to 20\%. For the sources where a preliminary estimate of the period was available, a period range of $\pm10\%$ around the estimated period was searched (see \autoref{sec:MeerTRAP}). Candidate periods with $\text{S/N}>8$ were used to fold the data for manual inspection. The folded profiles of detected sources are shown in \autoref{fig:subintegration}.

\section{Results}\label{sec:results}
\begin{figure*}
\centering
\includegraphics[width=0.9\textwidth]{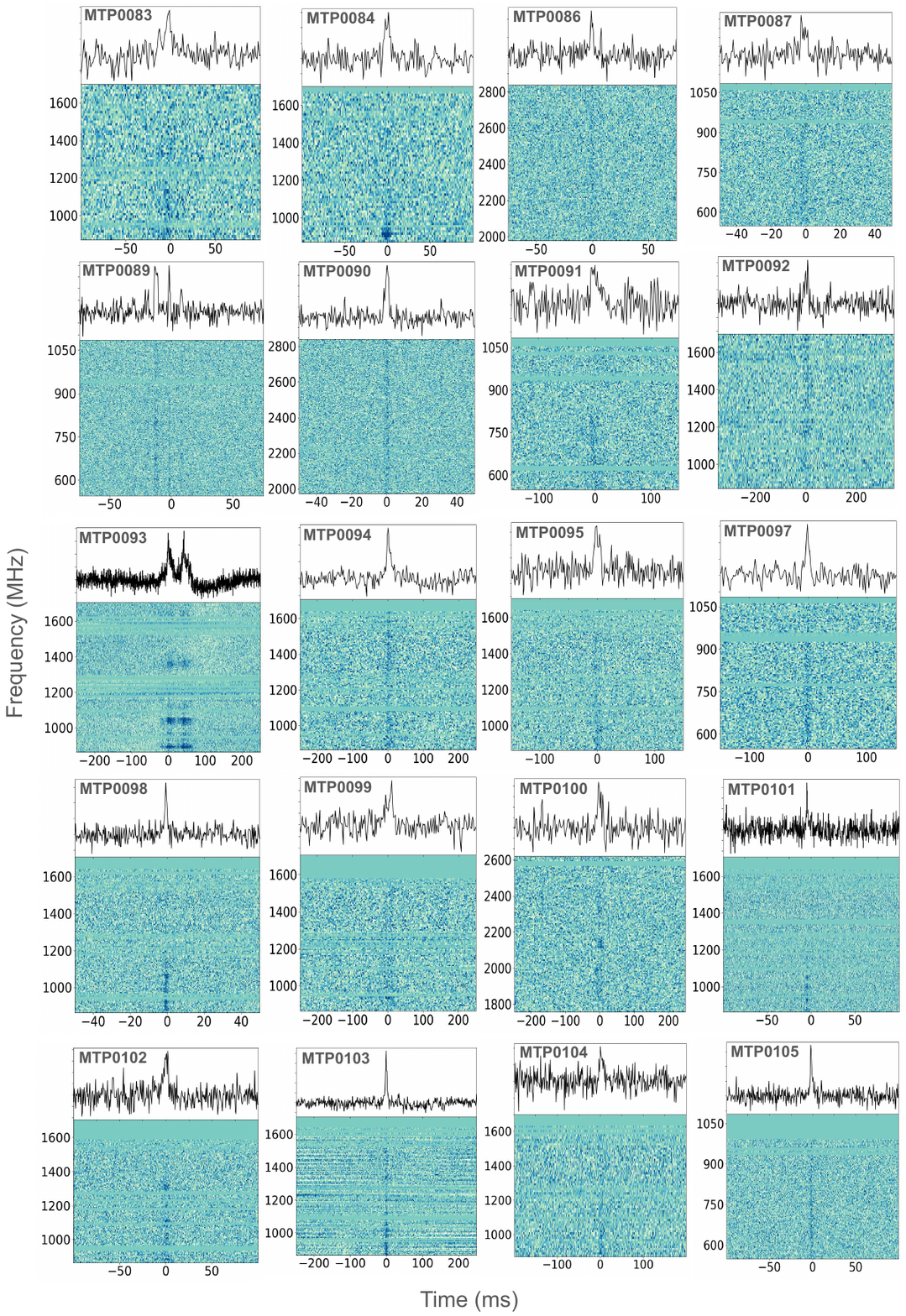}
\caption{Discovery plots of 27 new galactic transients. Each panel shows the dynamic spectra and the frequency-averaged pulse. The source names (see \autoref{tab:detections}) are indicated in the top left of each panel. The masked channels can be identified as horizontal lines in the dynamic spectra.}
\label{fig:discovery}
\end{figure*}

We report the discoveries of 27 new Galactic radio transients as part of the MeerTRAP programme. The discovery pulses along with dynamic spectra are shown in \autoref{fig:discovery}, and discovery parameters are listed in \autoref{tab:detections}. Out of the 27 discoveries, 15 were discovered at the L-band, 9 at the UHF band, and, for the first time, 3 at the S band. Six new sources were found in the IB, and the remaining 21 were found in the CB.
Additional properties of the discoveries, derived from MeerTRAP detections, such as period, refined DM, distances estimated using existing Galactic electron density models \citep{NE2025, NE2001, YMW16}, fluence, and location, are listed in \autoref{tab:properties}. These discoveries, along with discoveries reported in \citet{Bezuidenhout22, MeerTRAP_james, MeerTRAPIII}, bring the total number of Galactic radio transients discovered by MeerTRAP to 95.

\begin{table*}
\centering
\begin{tabular}{llcclllrrrrl}
\hline
MTP name & PSR J2000 & Discov. MJD & Discov. project$^{1}$ & Discov. & Discov. &  Discov. DM & $T_{\text{obs}}^{3}$ & $N_{\text{det}}^{4}$ & $N_{\text{ep}}^{5}$ & Burst rate$^{6}$ & Band$^{7}$ \\
&name&&&mode$^{2}$&S/N& (pc cm$^{-3}$)&(hr)&(hr)& & ($\text{hr}^{-1}$) &\\
\hline
MTP0083 & J1726$-$6456 & 60031.944278 & SCI-20220822-IP-01 & CB & 9.3 & 77.0 & 19.37 & 1229 & 8 & 63 & L\\
MTP0084 &  & 60013.892389 & SCI-20220822-FD-02 & CB & 9.4 & 24.2 & 1.05 & 4 & 1 & 3.8 & L\\
MTP0086 & J1811$-$1740 & 60075.059636 & SCI-20180516-MB-03 & CB & 9.0 & 398.0 & 10.77 & 43 & 9 & 4 & S, L\\
MTP0087 &  & 60064.895569 & SCI-20220822-WO-01 & CB & 10.6 & 205.0 & 0.18 & 16 & 1 & 88 & U\\
MTP0089 &  & 60116.961804 & SCI-20180516-MB-03 & CB & 14.3 & 21.4 & 0.16 & 1 & 1 & & U\\
MTP0090 & J1012$-$5757 & 60107.779447 & SCI-20200703-MK-03 & CB & 15.0 & 188.0 & 0.35 & 1 & 1 & & S\\
MTP0091 & J0937$-$6344 & 60223.313214 & SCI-20180923-MK-02 & IB & 8.6 & 133.8 & 0.53 & 1 & 1 & & U\\
MTP0092 &  & 60227.619262 & SCI-20180516-MB-02 & CB & 9.0 & 44.1 & 0.63 & 3 & 3 & 4.8 & L\\
MTP0093 &  & 60247.895264 & SCI-20180516-EB-01 & IB & 47.0 & 34.5 & 23.44 & 1 & 1 & & L\\
MTP0094 & J1449$-$5953 & 60297.562887 & SCI-20180923-MK-03 & IB & 13.3 & 307.0 & 28.79 & 3 & 2 & 0.1 & L\\
MTP0095 & J1454$-$3338 & 60301.471744 & SCI-20180516-MB-03 & CB & 10.4 & 51.0 & 31.11 & 14 & 6 & 0.45 & L, U\\
MTP0097 & J1826+1154 & 60125.768211 & SCI-20180923-MK-06 & CB & 11.4 & 78.0 & 0.31 & 1 & 1 & & U \\
MTP0098 & J1155$-$5021 & 60399.031560 & DDT-20240327-JL-01 & CB & 12.8 & 118.5 & 1.37 & 6 & 1 & 4.4 & L\\
MTP0099 &  & 60365.301896 & SCI-20230907-FS-01 & CB & 8.4 & 194.0 & 0.19 & 1 & 1 & & L\\
MTP0100 &  & 60360.289618 & DDT-20240213-AW-01 & CB & 10.8 & 230.0 & 7.62 & 1 & 1 & & S\\
MTP0101 & J0506$-$5900 & 60447.355664 & SCI-20230907-HB-01 & CB & 9.8 & 41.7 & 2.90 & 1 & 1 & & L\\
MTP0102 &  & 60363.155066 & SCI-20230907-MC-02 & IB & 11.8 & 36.2 & 9.73 & 2 & 1 & 0.2 & L\\
MTP0103 &  & 60535.578465 & SCI-20230907-LD-01 & IB & 14.7 & 40.5 & 30.82 & 1 & 1 & & L\\
MTP0104 &  & 60538.665171 & SCI-20230907-MC-02 & CB & 8.4 & 207.5 & 0.03 & 1 & 1 & & L\\
MTP0105 &  & 60586.625152 & SCI-20180923-MK-06 & CB & 14.5 & 117.1 & 0.50 & 1 & 1 & & U\\
MTP0106 & J1324$-$6145 & 60517.552349 & SCI-20230907-MC-02 & CB & 8.5 & 133.4 & 0.03 & 1 & 1 & & L\\
MTP0107 & J2225+1243 & 60623.772528 & SCI-20180516-NG-02 & CB & 10.6 & 45.0 & 1.48 & 114 & 1 & 77 & U\\
MTP0108 & J1410$-$4804 & 60729.232748 & SCI-20241101-KO-02 & CB & 36.1 & 63.8 & 0.73 & 1 & 1 & & L\\
MTP0109 &  & 60755.321620 & SCI-20180516-MB-05 & CB & 10.0 & 165.0 & 1.90 & 3 & 3 & 1.6 & U\\
MTP0110 & J1750$-$4450 & 60774.967866 & SCI-20180923-MK-06 & CB & 12.7 & 131.4 & 0.73 & 36 & 4 & 49 & U\\
MTP0111 & J0929$-$1855 & 60789.655258 & SCI-20250221-MB-01 & IB & 11.5 & 51.9 & 16.12 & 1 & 1 & & U\\
MTP0112 &  & 60793.797061 & SCI-20241103-MM-01 & CB & 9.3 & 106.5 & 5.41 & 3 & 1 & 0.6 & L\\
\hline
\end{tabular}
\caption{Discovery information for the reported transient sources. See \autoref{sec:results} for more details. \\\hspace{\textwidth}1: The proposal id of the discovery project for each source.\\\hspace{\textwidth}2: The observing mode of the discovery observation for each source with "CB" and "IB" denoting coherent and incoherent beam, respectively.\\\hspace{\textwidth}3: The total observing time in the discovery observation mode (IB/CB).\\\hspace{\textwidth}4: The total number of pulses detected from each source.\\\hspace{\textwidth}5: The number of epochs when the source was detected.\\\hspace{\textwidth}6: Burst rate for sources with more than one detection.\\\hspace{\textwidth}7: The observing band at which each source has been detected.}
\label{tab:detections}
\end{table*}

\begin{table*}
\centering
\begin{tabular}{lllllllll}
\hline
MTP name & RA & Dec & Localization & DM & Distance (kpc) & Fluence & $P$ &  P Epoch \\
&(hh:mm:ss)&(dd:mm:ss)& method & (pc cm$^{-3}$)& (\texttt{NE01/NE25/YMW16}) & (Jy ms) & (s)&(MJD)\\
\hline
MTP0083 & 17:26:18.5(1) & -64:56:33.0(12) & Imaging & 75(1) & 2.2/3.0/4.6  & 0.56 - 0.96 & 1.65428(4) & 60031.944278 \\
MTP0084 & 12:22:01.79 & 13:39:14 & CB & 24.1(9) & 1.6/3.1/25.0 & $\ge$ 0.17 - 0.33 & 2.60344(8) & 60013.892389 \\
MTP0086 & 18:11:30.9(1) & -17:40:52.9(15) & Imaging & 399(3) & 5.2/5.4/4.2 & 0.13 - 0.31 & 1.806466(2) & 60075.059636 \\
MTP0087 & 08:15:29 & -35:22:14 & CB & 205.1(2) & 3.4/3.8/3.6 & $\ge$ 0.38 - 1.48 & 1.54348(3) & 60064.895569 \\
MTP0089 & 19:45:48.4 & +21:28:43 & CB & 21.47(5) & 1.8/1.8/1.3 & $\ge$ 0.38 & & \\
MTP0090 & 10:12:57.9(1) & -57:57:44.4(12) & Imaging & 185.9(6) & 3.7/3.5/2.2 & 0.39 & &  \\
MTP0091 & 09:37:11.6(1) & -63:44:00.6(16) & Imaging & 134.3(4) & 4.0/5.4/2.0 & 11.5 & & \\
MTP0092 & 18:38:33 & -12:42:28 & CB & 42(2) & 1.3/1.4/1.0 & $\ge$ 0.52 & & \\
MTP0093 & 01:37:41 & +33:09:35.1 & IB & 34.9(7) & 1.6/2.6/2.8 & \ignore{32.2} & & \\
MTP0094 & 14:49:21.6(1) & -59:53:19.9(18) & Imaging & 307.1(7) & 4.8/4.7/5.0 & 3.18 & & \\
MTP0095 & 14:54:56.4(1) & -33:38:10.7(19) & Imaging & 51.5(4) & 1.6/2.3/3.4 & 0.25 - 0.48 & 2.9407918(3) & 60301.510070 \\
MTP0097 & 18:26:22.8(3) & +11:54:36.7(33) & Imaging & 78.0(4) & 3.2/4.7/4.9 & 0.65 & & \\
MTP0098 & 11:55:55.6(1) & -50:21:10.1(12) & Imaging & 118.3(1) & 3.7/5.5/2.2 & 0.19 - 0.38 &  0.7823670(4) & 60399.031560 \\
MTP0099 & 16:46:42 & -54:07:18 & CB & 192(1) & 4.3/5.5/8.9 & $\ge$ 0.46 & & \\
MTP0100 & 18:33:09.61 & -09:03:42 & CB & 224(6) & 4.2/4.5/3.6 & $\ge$ 0.88 & & \\
MTP0101 & 05:06:48.9(2) & -59:00:05.8(14) & Imaging & 42.8(3) & 2.7/5.0/25.0 & 0.29 & & \\
MTP0102 & 18:12:28.60 & -35:51:33.0 & IB & 36.7(3) & 1.0/2.5/1.0 & &  & \\
MTP0103 & 09:44:28.7 & -45:47:50 & IB & 40.6(1) & 0.5/0.5/0.1 & \ignore{3.18} & & \\
MTP0104 & 17:07:06.93 & -50:48:09 & CB & 206(1) & 4.8/6.1/12.5 & $\ge$ 0.31 & & \\
MTP0105 & 16:39:59.34 & -27:34:59 & CB & 117.21(7) & 3.5/4.9/15.2 & $\ge$ 0.33 & & \\
MTP0106 & 13:24:00.8(1) & -61:45:14.6(7) & Imaging & 134(1) & 2.5/2.3/2.4 & 0.64 & & \\
MTP0107 & 22:25:10.2(1) & +12:43:17.2(6) & Imaging & 45.1(1) & 41.9/12.9/25.0 & 0.54 - 1.57 & 4.007728(8) & 60623.772528\\
MTP0108 & 14:10:20.7(1) & -48:04:37.0(3) & Imaging & 63.84(4) & 1.7/2.0/2.0 & 1.69 &  & \\
MTP0109 & 18:01:31 & -14:41:38 & CB & 165.5(6) & 3.5/4.0/4.3 & $\ge$ 0.64 & & \\
MTP0110 & 17:50:20.4(1) & -44:50:25.8(5) & Imaging & 130.8(3) & 3.2/4.0/8.2 & 0.56 - 2.19 & 2.1951222(8) & 60774.967866 \\
MTP0111 & 09:29:14.1(1) & -18:55:23.9(3) & Imaging & 52.0(5) & 2.5/4.5/7.5 & 4.23 & & \\
MTP0112 & 10:43:13.94 & -46:09:02 & CB & 105.7(7) & 3.6/5.7/2.2 & $\ge$ 0.18 & & \\
\hline
\end{tabular}
\caption{Position, Localization method, DM, distance, fluence, and period measured for each transient source. We provide positional uncertainties for the sources localised with imaging. For the other sources, their positional errors can be estimated by the beam size of the respective observing mode. The mean beam sizes in the UHF, L, and S bands are 76$^{''}$, 48 $^{''}$, 26$^{''}$ for coherent beams, and  93$^{'}$, 59$^{'}$, and 31$^{'}$ for incoherent beams, respectively. We provide a range of fluences measured for individual pulses for the sources with imaging localization and a lower limit for the other sources, as we do not know the source location in the beam. The periods estimated using {\sc rratsolve} are listed with uncertainties for sources with multiple pulse detections.
}
\label{tab:properties}
\end{table*}

\renewcommand{\thefigure}{\arabic{figure} (Continued.)}
\addtocounter{figure}{-1}
\begin{figure*}
\centering
\includegraphics[width=0.9\textwidth]{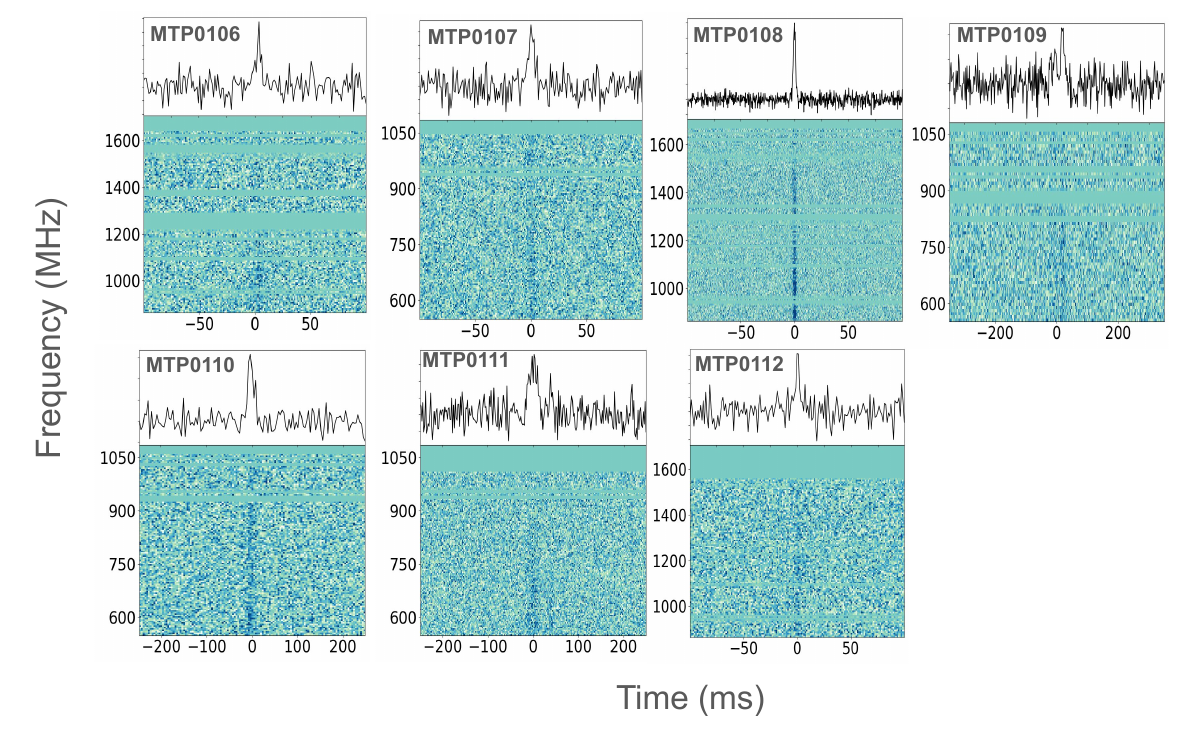}
\caption{}
\end{figure*}
\renewcommand{\thefigure}{\arabic{figure}}

\subsection{Sources independently discovered in other surveys}
Some of the sources reported in this work have been detected or discovered independently by other radio transient search programs. MeerTRAP discovered the source MTP0084 on 10 March 2023 at a DM of 24.1 pc cm$^{-3}$ at the L band. The Commensal Radio Astronomy FasT Survey (CRAFTS, \citealt{FAST_CRAFT_SETUP}) found the same source in a single pulse search at a DM of 25.7 pc cm$^{-3}$ on 30 August 2024 \footnote{\url{http://groups.bao.ac.cn/ism/CRAFTS/202203/t20220310_683697.html}}. Similarly, MTP0106 was discovered by MeerTRAP on 26 July 2024 at a DM of 133.4 pc cm$^{-3}$ at the L band. We detected only a single pulse from this source and localized it. The MPIfR-MeerKAT UHF Galactic Plane Survey discovered the same source on 5 February 2026\footnote{\url{https://www.trapum.org/discoveries/}} in a periodicity search, with a period of 668 ms and a DM of 133.3 pc cm$^{-3}$ (Bhatnagar et al. 2026 in prep.). 
MTP0093 is one of the transients discovered in the IB at a DM of $\sim$35 pc cm$^{-3}$ and has an uncertainty of 1 degree in its location. Notably, a new pulsar, J0149+29 \citep{Pushchino24}, was discovered at a similar DM value of 34.5 pc cm$^{-3}$ with a positional uncertainty of  $\sim 20$ arcmin but is situated $\sim$ 5 degrees away from the IB center corresponding to the MTP0093 discovery. The 5 degrees separation means that J0149+29 will fall in the second sidelobe of the IB, where the instrument sensitivity is reduced by a factor of $\sim$1000 and becomes heavily frequency dependent \citep{MeerKAT_primarybeam}, making it very unlikely to be detected. Given the significant distance between the two sources and the fact that we detected only a single pulse from MTP0093, we can assume that MTP0093 is different from pulsar J0149+29 until further pulses are seen with MeerTRAP that can be used to check the period or location of the source. MTP0100 was discovered on 20 Feb 2024 by MeerTRAP in a CB. Only a wide single pulse ($\sim$14 ms, see \autoref{fig:discovery}) was detected at the S-band at a DM of 230 pc cm$^{-3}$. Recently, \citet{pulsar_gleaners} found a pulsar (J1832$-$0901t) close to the location of MTP0100 at a DM of 235.2 pc cm$^{-3}$. This pulsar has a period of 29.7 ms, and the on-pulse window is $\sim$5 ms (see Fig 2 of \citealt{pulsar_gleaners}). The size of the MeerTRAP CB at the S-band is 0.5 arcmin, and the positional uncertainty of the FAST beam is 1.5 arcmin, while the separation between the MTP0100 and pulsar J1832-0901t is around 6 arcmin. Given the difference between the on-pulse width of PSR J1832$-$0901t (5 ms at 1250 MHz) and the width of MTP0100 pulse (14 ms at 2405 MHz), and the difference in the sky coordinates, we consider MTP0100 to be different from PSR J1832$-$0901t.

\subsection{Localization and Timing}
\autoref{tab:properties} lists the sky coordinates for all 27 new discoveries. We were able to perform image domain localization for 14 out of the 27 discoveries using the transient buffer data, resulting in arcsecond accuracy. We provide the uncertainty in Right Ascension (RA) and Declination (Dec) in \autoref{tab:properties} for the sources localized in the image domain. Image domain localization was also done for MTP0040, a MeerTRAP discovery reported in \citet{MeerTRAP_james}. The position of MTP0040 was determined to be \text{RA} = 13:57:55.8$\pm$1.3\,{\text{s}} and \text{Dec} = -65:07:18.8$\pm1.6^{\prime\prime}$. We also determined the period of MTP0040, now PSR J1357$-$6507 after localization, to be 1.8744(4) s using the recent detections. The uncertainty in the position of sources that were not localized in imaging is equivalent to the size of the beams they were discovered in. For the sources discovered in the CB, the uncertainty is $\sim 1$ arcmin, while for the IB discoveries, the uncertainty is $\sim 1$ degree.

\begin{figure*}
    \subfigure[PSR J1726$-$6456]{\includegraphics[width=0.8\linewidth]{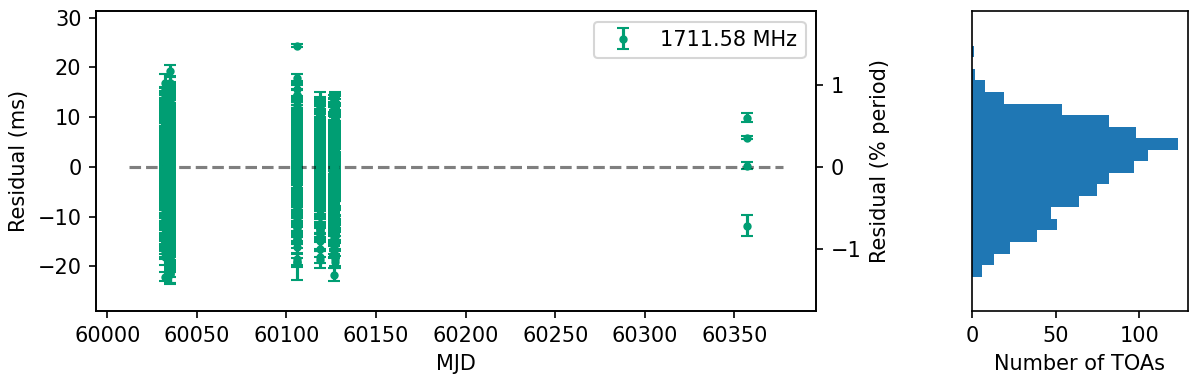}}
    \subfigure[PSR J1811$-$1740]{\includegraphics[width=0.8\linewidth]{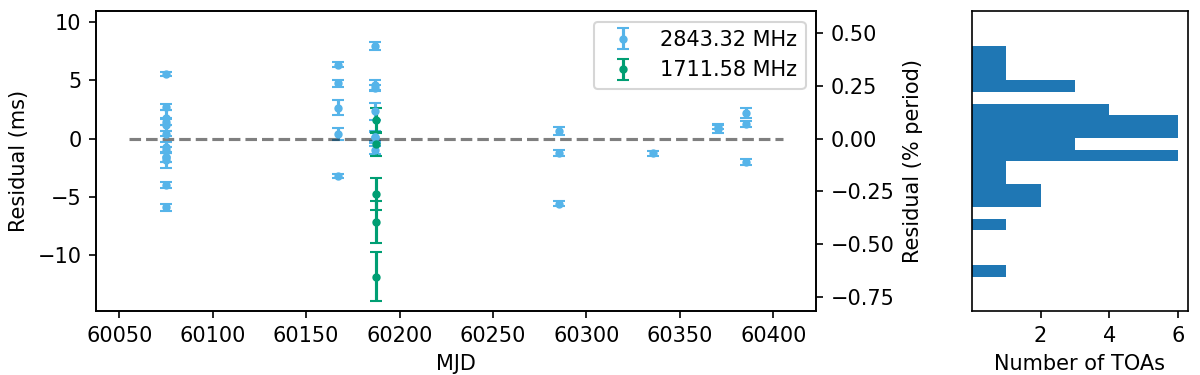}}
    \subfigure[PSR J1454$-$3338]{\includegraphics[width=0.8\linewidth]{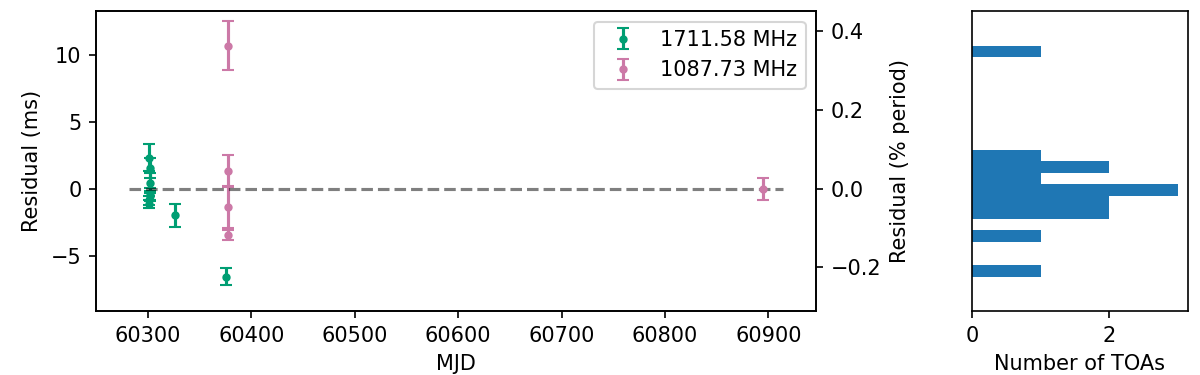}}
    \subfigure[PSR J1750$-$4450]{\includegraphics[width=0.8\linewidth]{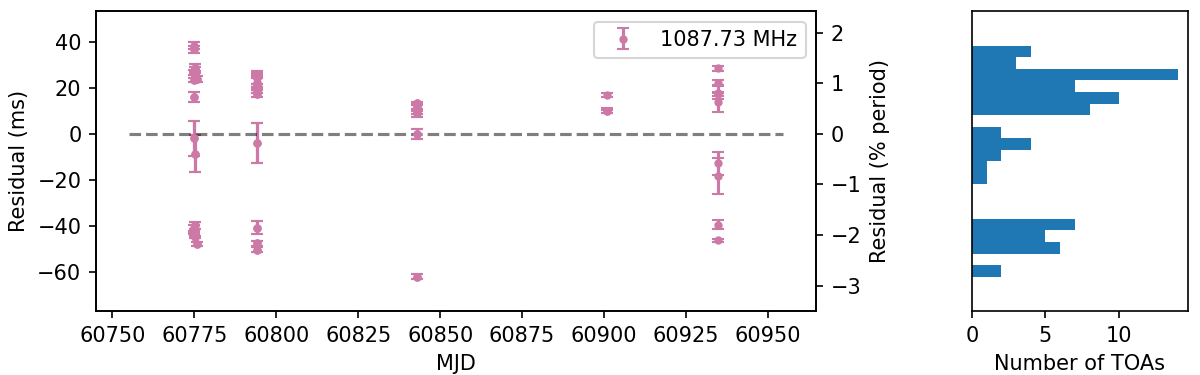}}
    \caption{Best fit timing residual for four discoveries. The right panel shows the histogram of the residuals. We could constrain the period derivative for the first three sources.}
    \label{fig:Timing}
\end{figure*}

We obtained phase-coherent timing solutions for some of the image-localized sources detected on multiple days across several months. We used the coordinates from the image localization and an initial period estimated by running {\sc rratsolve}\footnote{\url{https://github.com/v-morello/rratsolve}} on several pulses detected on a single day to construct an initial timing parameter file. Then, we used {\sc tempo2} \citep{hobbs2006} with the initial parameter file and TOAs generated from all detections of the target source and fitted for period and period derivative. \autoref{fig:Timing} shows the timing residuals, and \autoref{tab:parfile} shows the parameters obtained from fitting for four discoveries: PSR J1726$-$6456, PSR J1811$-$1740, PSR J1454$-$3338, and PSR J1750$-$4450.

PSR J1726$-$6456 was detected on eight different days between 28 March 2023 and 17 February 2024, providing approximately one year of timing baseline. With this baseline and TOA distribution, we were able to constrain both the period and the period derivative of this source. The timing residuals show a single-peaked distribution (see \autoref{fig:Timing}(a)), indicating a single-component profile, consistent with the folded profile of this source shown in \autoref{fig:subintegration}(a) and discussed in \autoref{sec:folding}. We had 39 detections of PSR J1811$-$1740 on eight different days, providing a 310-day-long timing baseline. The timing residuals after fitting are shown in \autoref{fig:Timing}(b). The residuals follow a narrow single-peaked distribution, indicating a narrow single-component profile, consistent with \autoref{fig:subintegration}(b). For PSR J1454$-$3338, we only had 14 detections; the first 13 pulses were detected within a span of three months, and the last detection was made after 517 days. All the pulses on the first three epochs were detected in the L band, and the last two epochs had only UHF band detections. We first fitted for period and period derivative using the three L-band epochs. After fitting these three L-band TOA clusters, the remaining UHF band TOAs lined up within $1\%$ of the period, indicating a high chance of phase coherence. Then we refitted using all the TOAs, constraining the period and period derivative, and the timing residuals are shown in \autoref{fig:Timing}(c). For PSR J1750$-$4450, we had a short timing baseline of only 160 days and could not constrain the period derivative. The timing residuals for this source are shown in \autoref{fig:Timing}(d), where one can clearly see a bimodal distribution of timing residuals, indicating two components in the folded profile of this source.

\begin{table*}
\centering
\caption{Timing parameters for 4 discoveries.}
\label{tab:parfile}
\begin{tabular}{lllll}
\hline\hline
\multicolumn{5}{c}{Fit and data-set} \\
\hline
PSR name\dotfill & J1726$-$6456 & J1811$-$1740 & J1454$-$3338 & J1750$-$4450\\
MeerTRAP name\dotfill & MTP0083 & MTP0086 & MTP0095 & MTP0110 \\
MJD range\dotfill & 60031 -- 60357 & 60075 -- 60385 & 60301 -- 60894 & 60774 -- 60934 \\ 
Number of TOAs\dotfill & 989 & 39 & 14 & 76 \\
rms timing residual (ms)\dotfill & 7.4 & 3.4 & 1.6 & 29.7 \\
Weighted fit\dotfill &  Y   & Y & Y & N \\ 
Reduced-$\chi^{2}$\dotfill & 66.5 & 132.9 & 11.3 & 985 \\ 
\hline
\multicolumn{5}{c}{Measured Quantities} \\ 
\hline
Right ascension, $\alpha$ (hh:mm:ss)\dotfill & 17:26:18.5(1) & 18:11:30.9(1) & 14:54:56.4(1) & 17:50:20.4(1) \\ 
Declination, $\delta$ (dd:mm:ss)\dotfill & $-$64:56:33.0(1) & -17:40:52.9(15) & -33:38:10.7(19) & -44:50:25.8(5)\\ 
Spin period, $P$ (s)\dotfill & 1.6543954814(2) & 1.8065860258(4) & 2.9409767754(6) & 2.195311081(6) \\ 
First derivative of $P$, $\dot{P}$ ($\times 10^{-15}$\,ss$^{-1}$)\dotfill & 2.94(2) & 159.89(3) & 2.46(2) & 1.2(9)\\
Epoch (MJD)\dotfill & 60031.9 &  60075.05 & 60301.5 & 60774.9\\ 
Dispersion measure, DM (cm$^{-3}$pc)\dotfill & 75(1) & 399(3) & 51.5(4) & 130.8(3)\\ 
\hline
\multicolumn{5}{c}{Models} \\
\hline
& Clock correction procedure\dotfill & TT(TAI)  \\
& Solar system ephemeris model\dotfill & DE405 \\
& Binary model\dotfill & NONE \\
& Model version number\dotfill & 5.00 \\ 
\hline
\end{tabular}
\end{table*}

\begin{figure}
    \centering
    \includegraphics[width=\linewidth]{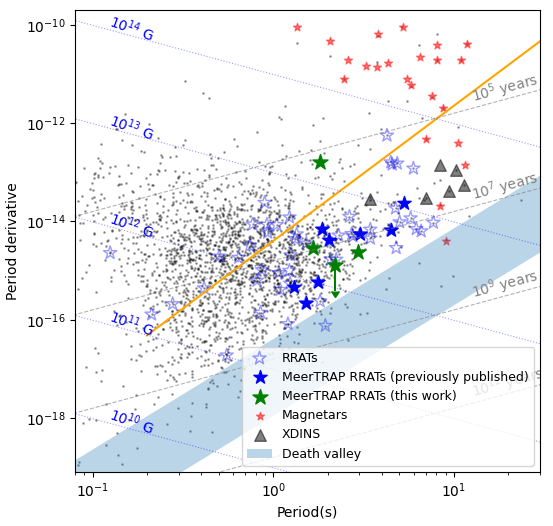}
    \caption{Normal pulsar population along with RRATs and magnetars on the $\text{P}-\dot{\text{P}}$ plane. The RRATs discovered with the MeerTRAP are shown as filled stars. The filled green stars show the position of four sources from this work. We were only able to obtain an upper limit on the period derivative for MTP0110 (PSR J1750$-$4450). The blue shaded region represents the vacuum voltage gap death valley as described by \citet{death_valley_emission}. The orange line has been adopted from \citet{pulsar_magnetosphere_philippov_kramer} and marks the transition from energetic young pulsars to the older pulsar population.}
    \label{fig:p_pdot}
\end{figure}

\subsection{Burst rate and Fluence}
We estimate the burst rate of the newly discovered sources based on the observation durations and the number of pulses detected with MeerTRAP. \autoref{tab:detections} lists the burst rates for the sources that were detected more than once in the MeerTRAP search. These estimates are lower limits on the actual burst rates, given that these sources were not targeted but happened to be covered by the MeerTRAP IB/CB, and the sensitivity of the search varies depending on the angular distance between the source and the beam center. Of the 27 sources reported in this work, 14 sources were detected only once. However, MTP0093 and MTP0103 were discovered in the IB and observed for 23.4 and 30.8 hours, respectively; despite this, they were detected only once, indicating highly intermittent/sporadic emission. The highest number of pulses was detected from PSR J1726$-$6456, 1229 pulses in 19.37 hours of CB observations. Six other transients were detected with 10 or more pulses. We see high burst rates of 63, 88, 77, and 49 $\text{hr}^{-1}$ from four new transients: PSR J1726$-$6456, MTP0087, PSR J2225+1243, and PSR J1750$-$4450, respectively. Incidentally, PSR J1726$-$6456 and PSR J2225+1243 were found to exhibit persistent pulsar-like emission in the follow-up observations, as discussed in \autoref{sec:folding}. PSR J1750$-$4450 was also followed up, but no persistent periodic signal was found in the periodicity search.

We calculate the radiometric fluences for the individual detections of the newly discovered sources (see \citealt{Tian24b, MeerTRAPIII} for details of the calculation method). The fluences of individual sources are listed in \autoref{tab:properties}. A range of fluence values has been provided for sources with more than one detection. Corrections were made for the response of the IB and CB. In the case of image-localized sources, we corrected for both the IB as well as the offset between the source location and the CB center\footnote{\url{https://github.com/BezuidenhoutMC/beam-corrections}}. We could only do IB correction for sources found in the CB, with no image localization. We only report the lower limit on fluences in such cases, as the correction for the offset between the actual location and the CB center is unknown. We do not report fluence values for sources found in an IB and with no image localization.

Most of the sources detected in the CB have a fluence value/limit of less than 1 Jy ms. Four CB discoveries, namely MTP0087, PSR J1410$-$4804, PSR J2225+1243, and PSR J1750$-$4450, show some bright pulses with fluence of more than 1 Jy ms. We include fluence measurements of 3 IB discoveries (PSR J0937$-$6344, PSR J1449$-$5953, and PSR J0929$-$1855) that were localized. These three sources show high fluences ranging between 3 and 11 Jy ms. Since they are localized, these sources form a good sample for follow-up studies with other telescopes. We also report the fluences of three S-band discoveries. Multiple pulses were detected from PSR J1811$-$1740 in the S-band, and the fluences range between 0.13 and 0.31 Jy ms for this source. Only one pulse from PSR J1012$-$5757 with a fluence of 0.39 Jy ms was detected in the S-band. The third S-band discovery, MTP0100, was found in a CB but could not be localized by imaging because the transient buffer was not triggered. The CB had been formed using only 16 MeerKAT dishes, and still MTP0100 was detected with a significant S/N of $\sim10$. The derived lower limit on the fluence of 0.88 Jy ms indicates that MTP0100 is a bright S-band transient.

\subsection{Persistent emission and folded profiles}\label{sec:folding}
The sources with a high burst rate and sufficient localization accuracy (sources found in CB or localized in the image domain) were followed up with the Effelsberg and Murriyang telescopes. We observed three MeerTRAP discoveries: PSR J1848+0009 \citep{MeerTRAPIII}, MTP0084, and MTP0089, with the Effelsberg telescope. We did not detect any signal in the single pulse search and in the FFA search for these sources, which was probably due to the high levels of RFI contamination suppressing any significant astrophysical signal.

We observed eight MeerTRAP discoveries (PSR J1726$-$6456, PSR J1811$-$1740, PSR J1155$-$5021, PSR J2225+1243, PSR J1410$-$4804, MTP0109, PSR J1750$-$4450, and PSR J0929$-$1855) with the Murriyang telescope. We detected four sources in the FFA search. \autoref{fig:subintegration} shows the folded profiles along with the time-phase plot for these four sources. The time-phase plots were made by folding the full observations into 256 sub-integrations. We performed flux calibration on these folded profiles using the flux calibrator PKS 1934$-$638 and following the method described in \citet{Parkes_fluxcal_Dai19}. All of these sources have regular periodic signals similar to radio pulsars. \autoref{tab:duty} lists the folded profile parameters: duty cycle, S/N, frequency range in which the source was visible, and phase-averaged flux values in the same frequency range. We could not perform flux calibration for PSR J2225+1243 due to low S/N and bad RFI conditions, and hence, no flux value is available for this source. PSR J1726$-$6456 shows signs of nulling, while PSR J2225+1243 was only visible in the first 40 minutes of the 2.5-hour-long observation. However, due to low S/N, we did not attempt to estimate the nulling fraction for these two sources. The other two sources, PSR J1811$-$1740 and PSR J1155$-$5021, show a persistent signal throughout these follow-up observations. PSR J1155$-$5021 also shows two components in its folded profile. It should be noted that these sources are on the fainter side of the pulsar population and were still discovered in a single-pulse search. The detection of pulsar-like emission from these sources reaffirms the complementary role of single pulse searches and periodicity searches in pulsar surveys and highlights the importance of sensitive commensal single pulse searches.

\subsection{Bright single pulses from PSR J1410$-$4804}

\begin{figure*}
\subfigure{
  \includegraphics[width=.4\linewidth]{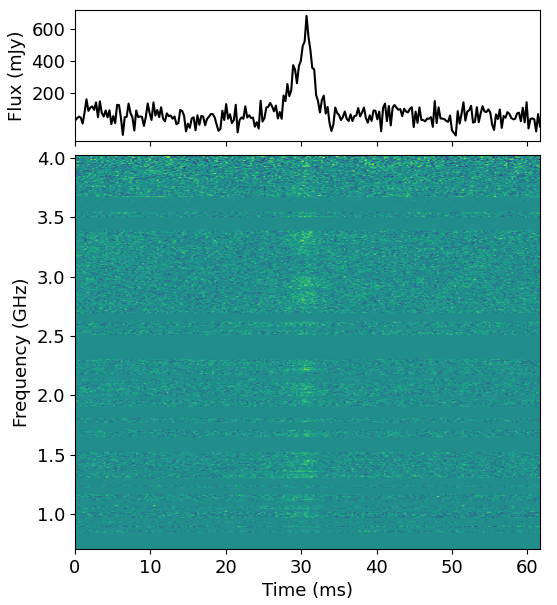}}
\subfigure{
  \includegraphics[width=.4\linewidth]{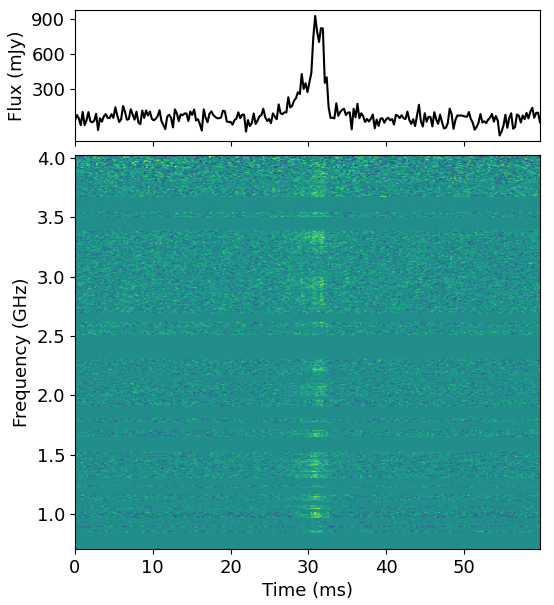}} 
\caption{Detection of two bright pulses from PSR J1410$-$4804 in the Murriyang follow-up observation. The upper panels show the shape of the pulses, while the lower panels show the de-dispersed dynamic spectra. The empty horizontal lines represent the masked frequency channels.}
\label{fig:MTP0108}
\end{figure*}

\begin{figure*}
\subfigure[PSR J1726$-$6456]{
  \includegraphics[width=.21\linewidth]{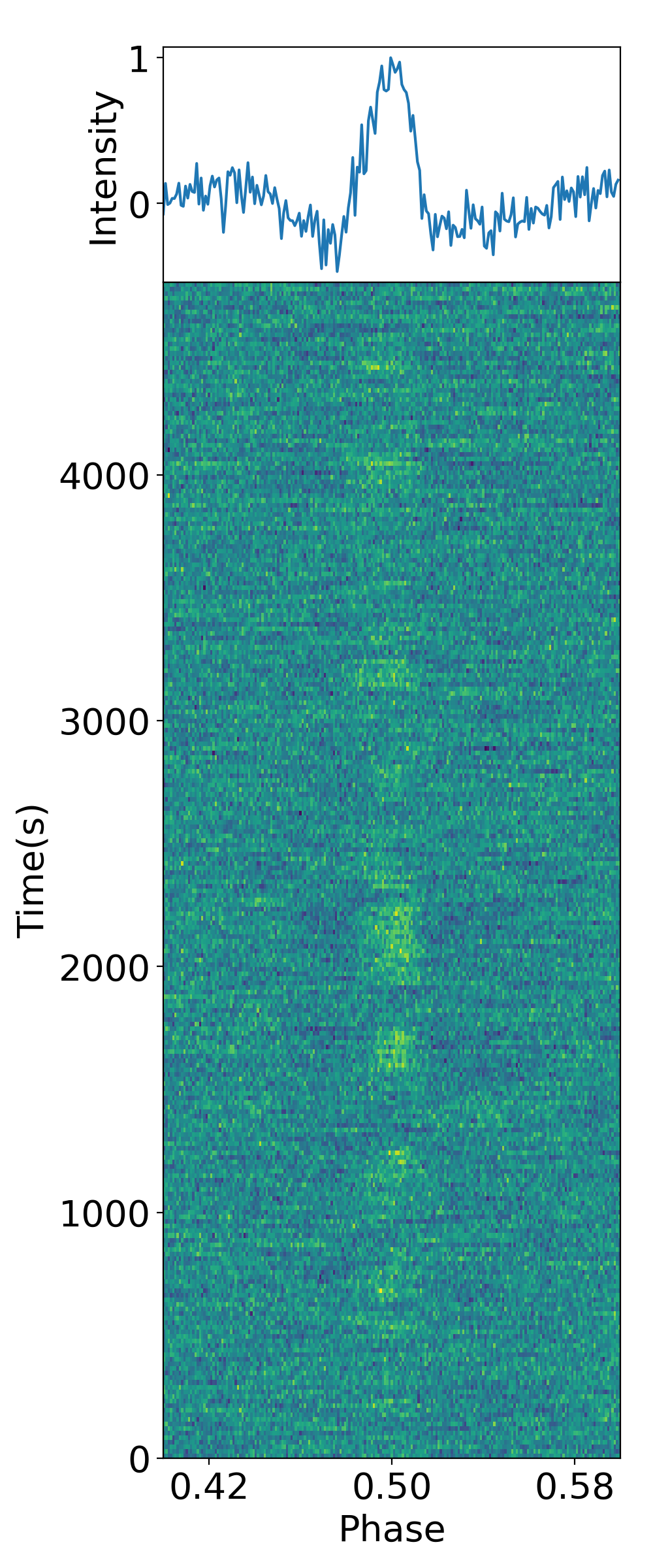}}
\subfigure[PSR J1811$-$1740]{
  \includegraphics[width=.21\linewidth]{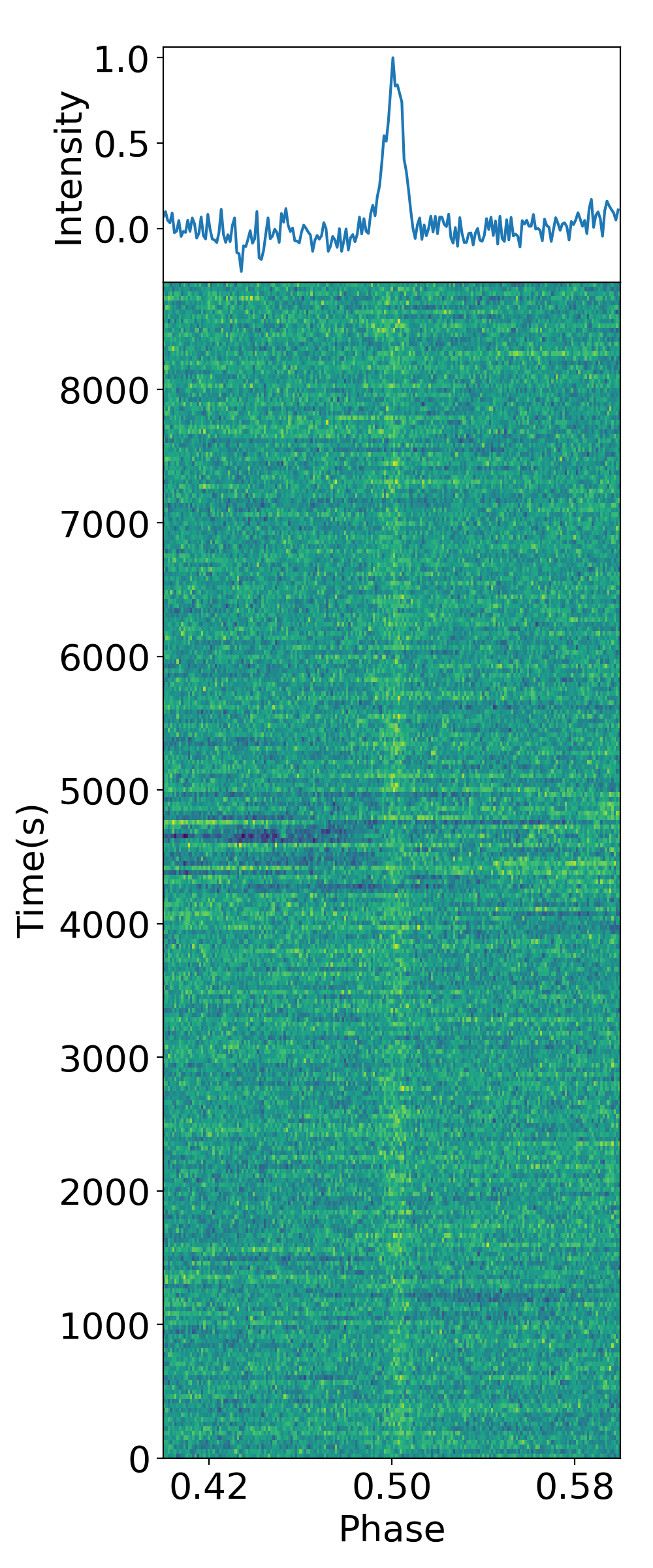}}
\subfigure[PSR J1155$-$5021]{
  \includegraphics[width=.21\linewidth]{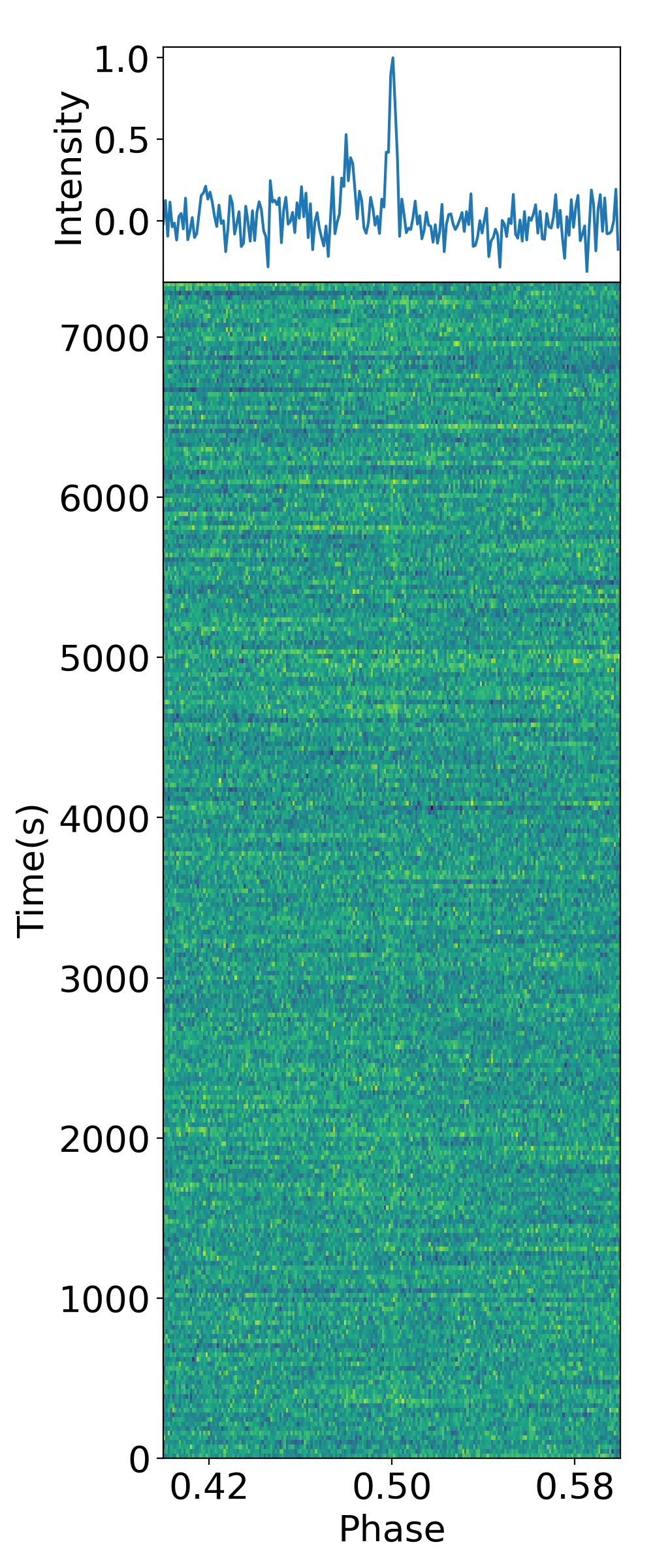}}
\subfigure[PSR J2225+1243]{
  \includegraphics[width=.21\linewidth]{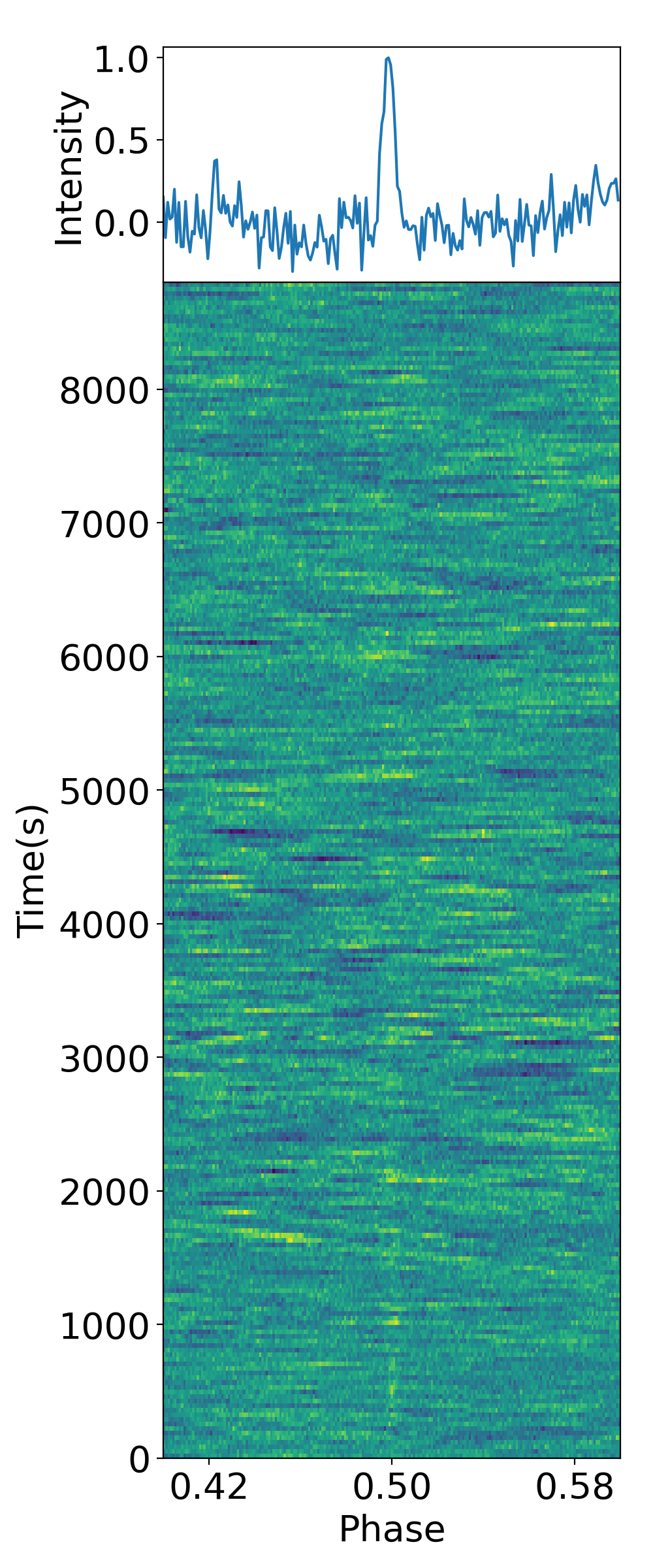}}  
\caption{Folded profiles and time-phase plots of four new MeerTRAP sources from Murriyang follow-up observations. While two sources, PSR J1811$-$1740 and PSR J1155$-$5021, were visible in the full UWL frequency band (704 - 4032 MHz), the remaining two sources, PSR J1726$-$6456 and PSR J2225+1243, were visible only in a small part of the band. The plot shown for PSR J1726$-$6456 is made using the data in the frequency range 1120 - 1536 MHz, while the plot for PSR J2225+1243 utilizes data in the frequency range 704 - 1776 MHz.}
\label{fig:subintegration}
\end{figure*}

While no periodic signal was identified in the Murriyang observation of PSR J1410$-$4804, we found two bright single pulses with $\text{S/N}=38$ and 55, respectively, as shown in \autoref{fig:MTP0108}. These pulses can be seen across the full UWL receiver bandwidth (704 MHz -- 4 GHz), indicating a relatively flat spectrum. We used the flux calibrator PKS B1934$-$638 and the method described in \citet{Parkes_fluxcal_Dai19} to obtain the peak flux values for these two pulses. The pulse detected with an S/N of 38 has a peak flux of 685$\pm$45 mJy, and the second pulse found with an S/N of 55 shows a peak flux of 925$\pm$45 mJy (see \autoref{fig:MTP0108}). The discovery pulse was detected with an S/N of 44 in a MeerTRAP CB. However, image localization reveals that the source was located at the edge of the CB, and the combined response of the IB and CB was at 35$\%$ of the maximum response. The flux density of the discovery pulse was estimated to be $\sim 800$ mJy using the radiometer equation, which is similar to the flux density of the pulses detected in Murriyang data. MeerTRAP searched around $\sim 45$ minutes of CB data and detected only a single pulse. The Murriyang follow-up observation was also 45 minutes long, and only two pulses were detected. This suggests a low burst rate of $\sim 2$ pulses per hour for this source. This burst rate estimate is not robust due to the small number of detected pulses and the sensitivity limitations. The non-detection of any faint periodic emission from this source places it in the category of prototype RRATs. Also, the rare broadband bright pulses from this source are similar to the recent detection of two radio pulses from the XDINS (X-ray-dim isolated neutron star) candidate 2XMM J104608.7$-$594306 \citep{XDIN_radio, XDIN_radio_II}.

\begin{table*}
\centering
\caption{Source properties derived from the follow-up observations. Duty cycle measured at $50\%$ of the profile peak, folded profile S/N, part of the Murriyang UWL frequency coverage where signal was visible, and phase-averaged flux density in the same frequency range are listed for the four MeerTRAP sources shown in \autoref{fig:subintegration}.
}
\label{tab:duty}
\begin{tabular}{lccccc}
\hline
PSR Name & MTP Name & Duty cycle & S/N & Obs. frequency & Average flux density\\
   & & ($\%$) & & (MHz) & ($\mu$Jy)\\
\hline
J1726$-$6456 & MTP0083 & $(2.1\pm0.5)$ & 30.8 & 1120 - 1536 & 138$\pm$8\\
J1811$-$1740 & MTP0086 & $(1.0 \pm 0.1)$ & 37.3 & 704 - 4032 & 29$\pm$2\\
J1155$-$5021 & MTP0098 & $(2.1 \pm 0.2)$ & 12.5 & 704 - 4032 & 7$\pm$2\\
J2225+1243 & MTP0107 & $(0.7 \pm 0.2)$ & 17.7 & 704 - 1776 & $-$ \\

\hline
\end{tabular}
\end{table*}

\section{Discussion}\label{sec:disc}

The transients reported in this work include several interesting sources. Some sources, such as MTP0089, MTP0093, PSR J1750$-$4450, and PSR J0929$-$1855, show pulses with multiple components. The only pulse detected from source MTP0089 shows clear microstructure (see \autoref{fig:discovery}). MTP0093 shows a wide pulse ($\sim 100\,\text{ms}$) with two distinct components. Source PSR J1155-5021 shows a weak, persistent periodic emission in the follow-up observation and has two distinct components in its folded profile (see \autoref{fig:subintegration}). We detected many pulses from PSR J1750$-$4450, and some of these pulses have two components. The timing residual distribution of PSR J1750$-$4450 also confirms the two-component profile. PSR J1410$-$4804 has a low burst rate, and the detected pulses are very bright. We detected two bright pulses from this source in the follow-up observation with the Murriyang telescope, and both pulses show flat spectra and cover a very wide frequency range of 0.7 \text{--} 4 GHz (see \autoref{fig:MTP0108}).

The discoveries reported here also include sources like MTP0093, PSR J1449$-$5953, MTP0103, and PSR J0929$-$1855, which were observed for tens of hours, but still only a few pulses were detected in the MeerTRAP search. Such low burst rates indicate extremely sporadic and intermittent emission. Two of the discoveries, MTP0104 and PSR J1324$-$6145, were observed for a very short duration ($\sim 5$ minutes) and only one single pulse was found. Longer follow-up observations on these targets might reveal transients with good burst rates or transients with persistent emission. Two of the sources, PSR J1726$-$6456 and PSR J2225+1243, showing persistent periodic emission, also show signs of nulling. In particular, PSR J2225+1243 was not visible for more than half of the 2.4-hour follow-up observation, indicating the presence of long nulls in this source. This source is located away from the Galactic plane (Galactic latitude b = $-36.6^{\circ}$) and the minimum DM distance of 12.9 kpc was obtained using \texttt{NE2025} \citep{NE2025}, while \texttt{NE2001} \citep{NE2001} and \texttt{YWM16} \citep{YMW16} models return a DM distance of 41.9 and 25.0 kpc respectively, indicating a high chance of it being located in the Galactic halo. The vertical height of this source (z-distance) from the Galactic plane is 7.7 kpc as per NE2025. Considering a mean kick velocity in the vertical direction to be 244 km/s \citep{Galactic_psr_population2}, a pulsar born in the Galactic plane will take 30.9 million years to reach the z-distance of 7.7 kpc, indicating an approximate age of 30.9 million years for this source. The approximate age of 30.9 million years and a period of 4 seconds place this source close to the pulsar death valley on the $\text{P}-\dot{\text{P}}$ plane, making it an interesting target for timing analysis. For two other off-Galactic plane sources, MTP0084 and PSR J0506$-$5900, the \texttt{YMW16} model predicts a very high DM distance of 25 kpc, while \texttt{NE2001} and \texttt{NE2025} predict much smaller DM distances for these sources (see \autoref{tab:properties}). MTP0084, PSR J0506$-$5900, and PSR J2225+1243 are also the highest-DM pulsars within radii of 10, 6, and 9 degrees, respectively, making them useful for constraining Galactic electron density models in their respective directions.

 PSR J1811$-$1740, PSR J1012$-$5757, and MTP0100 are the first three MeerTRAP Galactic transients found with the high-frequency S-band receivers (1968\text{--}2843 MHz). The source PSR J1811$-$1740 was followed up with the Murriyang telescope, as MeerTRAP detected multiple pulses from this source in the L and S bands. We detected a regular pulsar-like periodic signal from this source in the follow-up observation (see PSR J1811$-$1740 in \autoref{fig:subintegration}). These discoveries highlight the potential of high-frequency transient surveys to detect high-DM transients, particularly those with shallower spectra.

\subsection{Regular periodic emission from single pulse discoveries}
The long follow-up observations of the MeerTRAP-discovered Galactic transients with other sensitive telescopes, such as the Effelsberg and Murriyang telescopes, offer deeper insight into their nature. Until now, we have followed up 10 transients with Murriyang and 8 transients with the Effelsberg telescope. Many of these sources show a regular periodic signal in the long follow-up observations, while some show nulling. This demonstrates that many RRAT-like objects are regular pulsars detected in the single-pulse search \citep{RRAT_as_PSR, RRAT_zhang24}. However, there were many sources for which we could not find any kind of faint regular periodic emission in the follow-up observations. Such sources might represent the canonical RRAT population \citep{RRAT_review} or the population of extremely sporadic, nulling, or intermittent pulsars. These follow-up studies show that the pulsar population is not yet saturated, and there are fainter pulsars that can be discovered in sensitive periodicity searches. The fact that we did not find low-level periodic emission from many of the MeerTRAP-discovered Galactic sources, even in the long follow-up observations, indicates that there are many more sporadic emitters that require a sensitive single-pulse search to be discovered.

\subsection{Rotating Radio Transients in the $\text{P}-\dot{\text{P}}$ diagram}
\autoref{fig:p_pdot} shows the population of RRATs along with the population of pulsars, magnetars, and XDINSs. The period and period derivative for the pulsars were taken from the ATNF pulsar catalogue\footnote{\url{https://www.atnf.csiro.au/research/pulsar/psrcat/}} \citep{ATNF}. The magnetar data points were obtained from the McGill Online Magnetar Catalogue\footnote{\url{https://www.physics.mcgill.ca/~pulsar/magnetar/main.html}}\citep{magnetarlog}, and data points for known RRATs were taken from the RRATalog\footnote{\url{https://rratalog.github.io/rratalog/}}. The periods and period derivatives for the XDINSs were collected from several timing papers \citep{XDIN_0420, XDIN_0720, XDIN_0806, XDIN_1308, XDIN_1856, XDIN_2143}. The timed RRATs from the MeerTRAP programme (\citealt{Bezuidenhout22, MeerTRAP_james} and this work) are also highlighted in the figure. We also include a contour adopted from \citet{pulsar_magnetosphere_philippov_kramer} to indicate the transition from an energetic young population to a less energetic older population. As noted by \citet{pulsar_magnetosphere_philippov_kramer}, the majority of the known RRAT population, including those discovered in MeerTRAP, are located below this line among the less energetic pulsar population.

A significant fraction of RRATs have relatively longer periods and overlap with the XDINSs and magnetars. These three manifestations of neutron stars have been proposed to represent different evolutionary stages of neutron stars. Thermal and magnetic field evolution suggests that magnetars and RRATs can become XDINSs after cooling and magnetic field decay \citep{NS_evolution2013, NS_evolution25}. As for the connection between magnetars and RRATs, they could be different subclasses of neutron stars born with different magnetic field strengths. Magnetars, XDINSs, and some RRATs \citep{Rea09, Archibald17} can be seen in X-rays. Both RRATs and magnetars are known to show sporadic radio emission, and the recent detection of two radio pulses from an XDINS candidate \citep{XDIN_radio, XDIN_radio_II} also strengthens the possibility of XDINSs representing a later evolutionary stage of magnetars and RRATs \citep{NS_evolution25, NS_evolution2013}.

A large fraction of the current RRAT population overlaps with the normal pulsar population (see \autoref{fig:p_pdot}). It has been seen that many of the seemingly RRAT-like sources can be classified as highly nulling pulsars or weak pulsars with occasional bright pulses \citep{FAST_RRATs}. Results from the long follow-up observations in this work also suggest the same (see \autoref{fig:Timing}). However, more than half of the sources we observed with the Murriyang and Effelsberg telescopes showed no signature of a faint periodic signal. Such sources might represent the extremely nulling pulsars or prototype RRATs \citep{FAST_RRATs}. There are also transients like J1410$-$4804 that show occasional bright pulses and no weak periodic emission. Since the current RRAT population is a mixture of weak pulsars, extremely nulling pulsars, and prototype RRATs, it is difficult to identify the region on the $\text{P}-\dot{\text{P}}$ plane preferred by the prototype RRAT population. A dedicated effort to separate the canonical RRAT population from the weak and nulling pulsars is needed to unveil the characteristics of the actual RRATs and their place in the diverse neutron star zoo.

\section{Conclusions}\label{sec:concl}
We have reported the discovery of 27 new Galactic transients from the MeerTRAP programme, bringing the total number of Galactic transients discovered in MeerTRAP to 95. The majority of the transients reported in this work can be categorized as RRATs with burst rates ranging between $\sim0.05$ and 88 per hour and fluence values ranging between 0.13 and 11 Jy ms. 

We provide sky coordinates of 14 discoveries with arcsecond accuracy and a preliminary estimate of the period for 8 of the reported discoveries. We also reported the findings from long follow-up observations of some of the new discoveries. Folded profiles with time-phase plots showing a continuous, regular periodic signal from four transients are presented. This regular pulsar-like emission from sources detected in single pulse searches further bridges the gap between RRATs and the canonical pulsar population. We were also able to obtain coherent timing solutions for 4 transients using the single pulse detections spread across several epochs and obtained period derivative values for 3 of them and an upper limit for the fourth one. These results highlight the importance of sensitive commensal surveys in uncovering the Galactic neutron star population with sporadic and intermittent radio emission.

\section*{Acknowledgements}
The MeerKAT telescope is operated by the South African Radio Astronomy Observatory (SARAO), which is a facility of the National Research Foundation, itself an agency of the Department of Science and Innovation. The authors thank the MeerKAT LSP teams for allowing commensal observing and the staff at SARAO for scheduling MeerKAT observations. MeerTRAP observations use the FBFUSE and TUSE computing clusters for data acquisition and storage. These instruments were designed, funded and installed by the Max-Planck Institut f{\"u}r Radioastronomie (MPIfR) and the Max-Planck-Gesellschaft. 
This publication includes data from observations with the 100-m telescope of the MPIfR at Effelsberg. The authors are thankful of Dr. Alex Kraus for scheduling the observations. We also acknowledge the insightful discussions on RRAT timing with Dr. Michael Keith.
This project has received funding from the European Research Council (ERC) under the European Union’s Horizon 2020 research and innovation programme (grant agreement no. 694745). JT and BWS acknowledge funding from an STFC Consolidated grant. 
JDT acknowledges support by Fondazione Cariplo/Cassa Depositi e Prestiti (Grant 2023-2560, PI: Papitto). 
M.C. acknowledges support of an Australian Research Council Discovery Early Career Research Award (project number DE220100819) funded by the Australian Government.
IPM further acknowledges funding from an NWO Rubicon Fellowship, project number 019.221EN.019. MB acknowledges support through the research grant ‘iPeska’ (PI: A. Possenti) funded under the INAF national call Prin-SKA/CTA approved with the Presidential Decree 70/2016.
For the purpose of open access, the author has applied a Creative Commons Attribution (CC BY) licence to any Author Accepted Manuscript version arising.
This research used version 2.2.0 of the ATNF Pulsar Catalogue, the SIMBAD data base, operated at CDS, Strasbourg, France \citep{Wenger2000}, and NASA’s Astrophysics Data System Bibliographic Services.

\section*{Data Availability}
The data underlying this article will be shared on reasonable request to the corresponding authors.


\bibliographystyle{mnras}
\bibliography{references, bibfile} 




\appendix


\bsp	
\label{lastpage}
\end{document}